# Signatures of quantum chaos in phonon-polariton billiards

Yinan Dong[1,2], Felix Liu[1], Ekrem Demirboga[3], Andrey Grankin[3], Dihao Sun[1], Yuchen Lin[1], Lukas Wehmeier[4], Matthew Fu[1], Cory R. Dean[1], Song Liu[5], James H. Edgar[5], Michael M. Folger[6], Victor M. Galitski[3], Dmitri N. Basov[1]

[1]Department of Physics, Columbia University, New York, NY, 10027, USA.

[2]Department of Applied Physics and Applied Mathematics, Columbia University, New York, NY, 10027, USA.

[3]Department of Physics, University of Maryland, College Park, MD, 20742, USA.

[4]Brookhaven National Laboratory, Upton, NY, 11973, USA.

[5]Department of Chemistry, Kansas State University, Manhattan, KS, 66506, USA.

[6]Department Physics, University of California San Diego, La Jolla, CA, 92093, USA.

## Abstract

*We use scanning near-field optical microscopy to image hyperbolic phonon polaritons in hexagonal boron nitride (hBN) billiards with integrable and chaotic geometries. In Sinai cavities, we observe irregular mode patterns consistent with quantum scarring, together with an unexpected sensitivity to weak probe perturbations. These random-wave features coexist with non-chaotic one-dimensional boundary modes arising from non-trivial polariton reflection at the billiard edge. As the billiard boundary becomes more complex, the Fourier transforms of the measured signals evolve toward a ring-like structure, consistent with Berry's random-wave conjecture. We develop a numerical framework based on the Helmholtz equation with generalized boundary conditions that encode angle-dependent reflection phase shifts. The calculated level statistics exhibit a crossover from Poisson-like behavior in integrable billiards to Wigner-Dyson-like behavior in chaotic geometries, with small deviations from the canonical form arising from the non-linear boundary conditions that require a self-consistent bulk-boundary analysis. Theoretical analysis of dissipative Green's function qualitatively reproduces the near-field data. These results establish mesoscopic van der Waals billiards as a rich platform for studying generalized chaotic dynamics of light-matter hybrid polaritons*.

## Main

Quantum mechanics is a wave theory, but generic wave problems are nontrivial. Outside a small set of separable models, the specific spectrum and eigenfunctions of the Helmholtz or Schrödinger operator are controlled by geometry, making billiards a canonical setting for studying wave dynamics. In this setting, integrable geometries support regular mode families and Poisson-like spectral statistics, whereas chaotic geometries exhibit irregular eigenfunctions and the Wigner-Dyson fluctuations associated with universal results from random-matrix theory (RMT)[1,2,3,4]. A central question is whether these universal signatures of "quantum chaos" survive in real wave platforms with loss and nontrivial boundary scattering. Recent experiments have captured signatures of chaos in quantum wave systems including photons[5,6,7,8] , electrons[9,10] and excitons[11]. However, the polaritonic chaos are still lacking experimental demonstration. Polaritons provide a distinct arena for quantum chaos because they are neither purely photonic nor purely material, but hybrid light-matter waves whose dynamics are shaped jointly by geometry and medium response.

In this work, we investigate experimentally and theoretically chaotic signatures of phonon polaritons hosted by hexagonal boron nitride (hBN). Polaritons in van der Waals materials[12] inherently bridge the electromagnetic waves with collective excitations of matter, providing high confinement, low-loss and strong light-matter interactions. Phonon polaritons in hBN[13,14] are coherent hybrid excitations of mid-infrared photons and longitudinal optical phonons. Their strong optical anisotropy gives rise to hyperbolicity enabling hBN as a versatile platform for exploring quantum wave phenomena such as hyperbolic super resonances[15] and superluminal singularities[16]. Here we exploit multiple orders of hyperbolic phonon polaritons to reveal signatures of chaotic polariton wave dynamics, including quantum scars, chaotic energy landscape, and Berry's conjecture of random waves, all of which are sensitive to the perturbations.

**Near-field imaging of hBN phonon polaritons**

We study quantum wave phenomena of hBN phonon polaritons in the mid infrared regime within the type-II Reststrahlen band[9,10]. As shown in Fig. 1a, the anisotropic optical phonon resonances drive opposite signs of the in-plane ($Re(\varepsilon_{//}) < 0$) and out-of-plane ($Re(\varepsilon_{\perp}) > 0$) dielectric constants yielding a hyperbolic isofrequency surface. Figure 1b demonstrates the calculated dispersion for natural hBN with a thickness of 72 nm as an example. Each branch corresponds to a distinct polariton order supported by vertical confinement. In Fig. 1c, we illustrate the experimental schematics for imaging multiple orders of hBN phonon polaritons using scattered-type scanning near field optical microscopy (s-SNOM). The light and dark blue dashed arrows (Fig. 1a and c) indicate the momentum ($k$) and Poynting vectors ($S$) of propagating polaritons with the solid arrows mark the 2D projection of the momentum ($k_{//}$) and energy flows ($S_p$). During s-SNOM imaging, a mid-infrared beam is focused by a parabolic mirror, producing a far field spot that exceeds the sample size. The metallic tip apex, with a radius of around 10 nm, provides sub-diffraction and background-free nearfield imaging with demodulation methods[17]. In our experiments, the metallic tip plays a dual role in both launching and collecting signals. First, it acts as an optical antenna that couples incident electromagnetic waves into near-field polaritons. After these polaritons propagate and undergo multiple reflections within confined geometries, the tip serves as a local scatterer, outcoupling the returning polariton field for signal collection. The interference between the launched polaritons and those reflected from the boundaries gives rise to standing-wave patterns.

Figures 1d and 1e present representative SNOM images of a semi-infinite sample and a confined sample with the same thickness (68 nm) and excitation energy (1429 cm-

[1]). The dominant periodic modulation observed in Fig. 1d originates from first-order phonon polaritons, whereas higher-order polaritons with smaller periods are more evident in confined shapes in Fig. 1e. These higher-order modes arise from a superposition of multiple higher order polariton branches (circles in Fig. 1b), resulting in a slightly flattened wave profile in Fig. 1e compared with Fig. 1d while the period is governed by the second order polariton (Fig. 1e). Owing to larger momentums and smaller volumes, the cavity enhancement[18,19] for higher order polaritons is stronger than first-order polaritons with details described in SI Sec.3.1. A sufficiently long propagation length is a prerequisite for the chaotic wave phenomena investigated in this work. As established in prior studies, the dispersion of free hBN phonon polaritons is tunable by the sample thickness[9], isotopic composition[20], and temperature[21]. For a determined dispersion, the free propagation length of first-order hBN phonon polaritons is determined by the quality factor $Q$ and polariton momentum by a simple relation $L_P = \frac{Q}{Re(k_{//})}$, which is tunable by the excitation energy. As calculated in Fig. 1f, the typical $L_P$ values with $Q = 30$ for different thicknesses of natural hBN range from several to tens of microns as a conservative estimation without cavity enhancement[18,19] and isotope effect[20]. These values are sufficient to support multiple round trips of polariton propagation within micron-scale billiards, satisfying the prerequisite for realizing chaotic billiard dynamics.

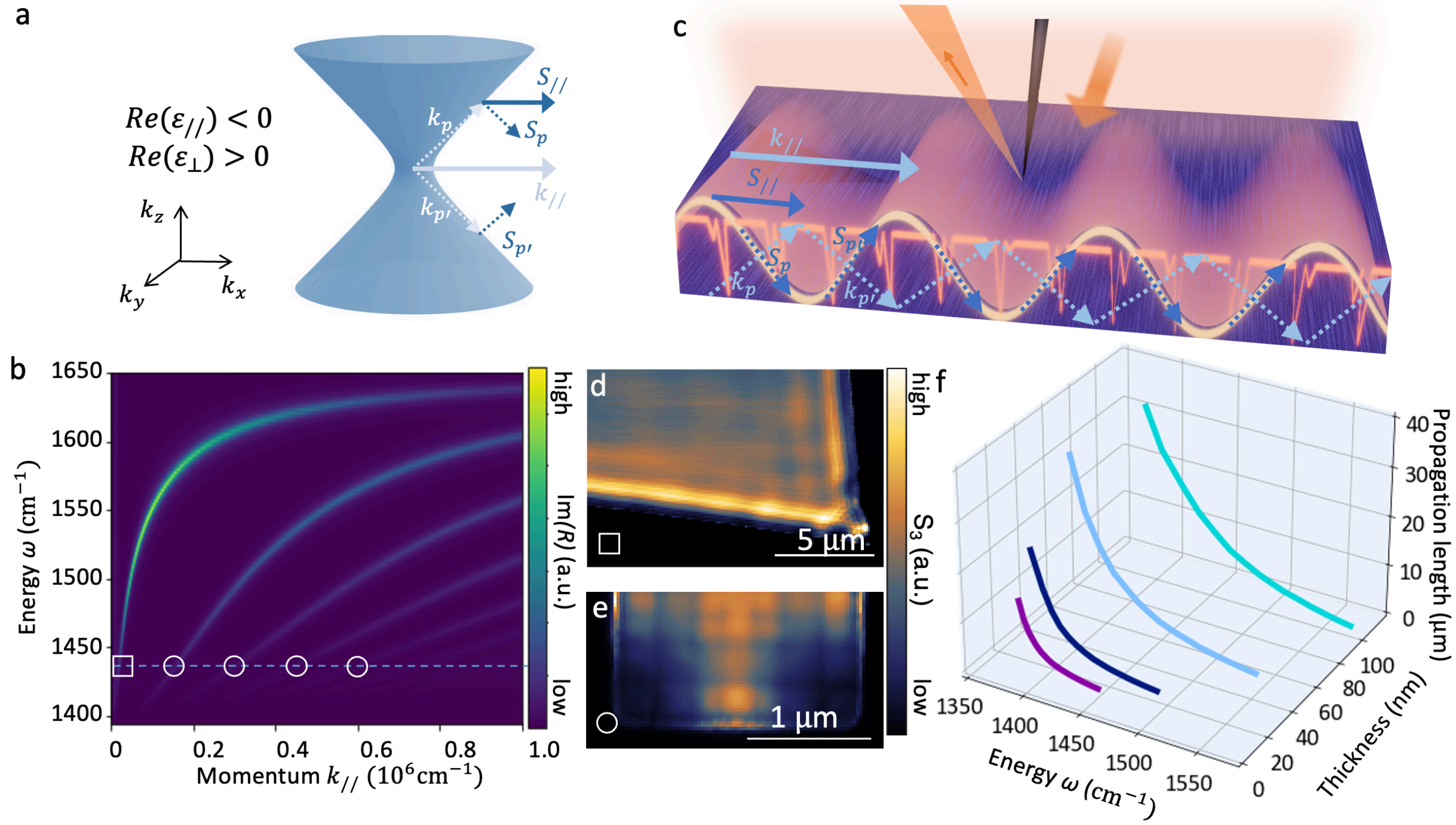

**Figure. 1 Nano-imaging of multiple orders of hBN phonon polaritons.** a. Illustration of type-II hyperbolic isofrequency surface with $Re(\varepsilon_{//}) < 0$ and $Re(\varepsilon_{\perp}) > 0$. The light blue dashed arrows show the momentum of polaritons that is defined by sample thickness and mode number. The dark blue dashed arrows indicate the corresponding Poynting vectors. The solid arrows mark the 2D projection of the net momentum and energy flow directions. b. Calculated dispersion of phonon polaritons for an $hB^{10}N$ slab of 68nm. The square and circles mark first order and higher orders of polariton modes with corresponding experimental data shown in d-e. c. Schematics of the experiment. The grey tip both launch and collect polariton fields. The first order polariton shows a sinusoidal shape while the higher order modes mix into a flatter shape. d. s-SNOM image of a corner of semi-infinite 68nm thick $hB^{10}N$ sample excited by laser with energy 1429 $cm^{-1}$ where first-order free-propagating polariton features dominate. e. s-SNOM image of the same material excited by the same laser energy but in a square cavity where higher-order polariton features dominate. f. Calculated thickness-dependent and energy-tunable first-order polariton propagation lengths in natural hBN with Q = 30.

## Signatures of chaos in a polariton Sinai billiard

The notion of polariton trajectory is poorly defined due to the uncertainty principle. Polaritons in a box (equivalent term to cavity or billiard) have quantized[22,23] momentum and energy while inheriting the intrinsic dispersion properties (SI Sec.3.2). By introducing scatterers or modifying boundaries, we can shape polariton billiards into non-integrable geometries and inquire the emerging chaotic phenomena. In Fig. 2, we focus on the paradigmatic chaotic geometry, Sinai billiard[24], to probe phonon polariton chaos. A Sinai billiard consist of a square domain and a central convex scatterer. Intuitively, the latter provides a negative curvature that defocuses the particles or waves propagating within the domain, and introduce chaoticity.

Scarring is a prominent phenomenon which captures the hidden ordering within chaotic quantum systems, where wave functions localize along unstable classical periodic orbits[25]. In polariton billiards, the scarring effect manifest itself as enhanced local density of states, or brightened traces in polariton maps, as demonstrated in Fig. 2a-i. Here we demonstrate three representative scar types, the cross (Fig. 2a-c), the diagonal (Fig. 2d-f) and the multi-bounce (Fig. 2g-i) in the first three columns in Fig. 2, respectively. The three rows are experimental SNOM data, Green's function simulations and classical orbit correspondence, respectively. SNOM data (third order pseudoheterodyne signals, $S_3$) in a and d are taken on a 7.6nm thick $hB^{10}N$ Sinai billiard at 1408.5$cm^{-1}$ and 1413 $cm^{-1}$; while data in I is taken on a 60nm thick natural hBN billiard at 1405$cm^{-1}$. All billiards are with 3 μm width and hole diameter 1.2μm. The cross-shape scar is the most common for Sinai billiards where polariton field is enhanced along the shortest distance. The diagonal scar[26] is also relatively common and can be observed in multiple scenarios (SI). The multibounce shape is observed only when higher order polaritons are dominant (Fig. 2g-i), which have much larger momentums and cavity enhancement (SI Sec. 3.1). In addition to mapping scarred modes, we note several unusual features observed in the SNOM imaging data. First, the darkened ring around the central perimeter in Fig. 2d reveal the decreased local density of states caused by defocusing effect, where edge launched modes are unable to interfere with the back-reflected path to form standing waves. Second, slightly unevenly spaced polariton wave front lines are observed (Fig. 2g) when higher order modes dominate. Furthermore, prominent edge modes[27] are observed along the square edges arising from dielectric discontinuities. The edge modes can be regarded as a separate, nearly integrable subsystem, while providing an additional dissipation channel for bulk modes. As detailed in SI Sec. 2.3, they naturally emerge when imposing the characteristic $\pi/4$ boundary reflection phase[30]. Additional data with different energies, billiard sizes and scatterer ratios can be found in SI. Overall, incorporating a finite quality factor into the dissipative Green's-function formalism yields excellent agreement with the experimental observations.

Beyond scarring, we further address the observed sensitivity to perturbations and the statistical evidence for polariton chaos. External measurement parameters such as probe amplitudes and signal demodulation orders are conventionally considered irrelevant to imaging of polariton patterns. In chaotic regimes, however, we have observed that polariton patterns can be altered by probe amplitudes and demodulation orders, as shown in Fig. 2 j-l and SI. At identical sample and excitation energy, Fig. 2 j and k show drastically different patterns from Fig. 2l with whispering gallery modes, by just tuning the probe amplitude. The perturbation from probe amplitude effectively modifies the initial condition for polariton launching, where boundary-induced phase accumulation and repeated reflections amplify small changes into global modifications of the near-field patterns. As we will discuss more in later section (Fig. 4 and SI), the experimentally observed sensitivity to perturbation reflects not only the local probe-sample coupling, but also the intrinsic geometry sensitivity of chaotic polariton dynamics. Another widely accepted signature of quantum chaos is the statistical distribution of energy levels. As expected from Bohigas- Giannoni-Schmit conjecture[3], the quantum systems whose classical limit are chaotic should possess statistical signature characterized by Wigner–Dyson statistics[28,29]. While our experiments with finite quality factors captures a mix of several energy levels, we can extract the energy level statistics by scrutinizing individual polariton eigenmodes that inform the simulations with nontrivial boundary conditions as required for matching experiments. As shown in Fig. 2m and n, the blue columns represent the normalized probability density of the nearest-neighbor energy level spacings for square and Sinai billiards, respectively. For integrable square billiards, the energy level spacings follow a Poisson distribution (dashed green curve); whereas energy level spacings of Sinai billiard polaritons conform to the Gaussian orthogonal ensemble (GOE, solid red curve).

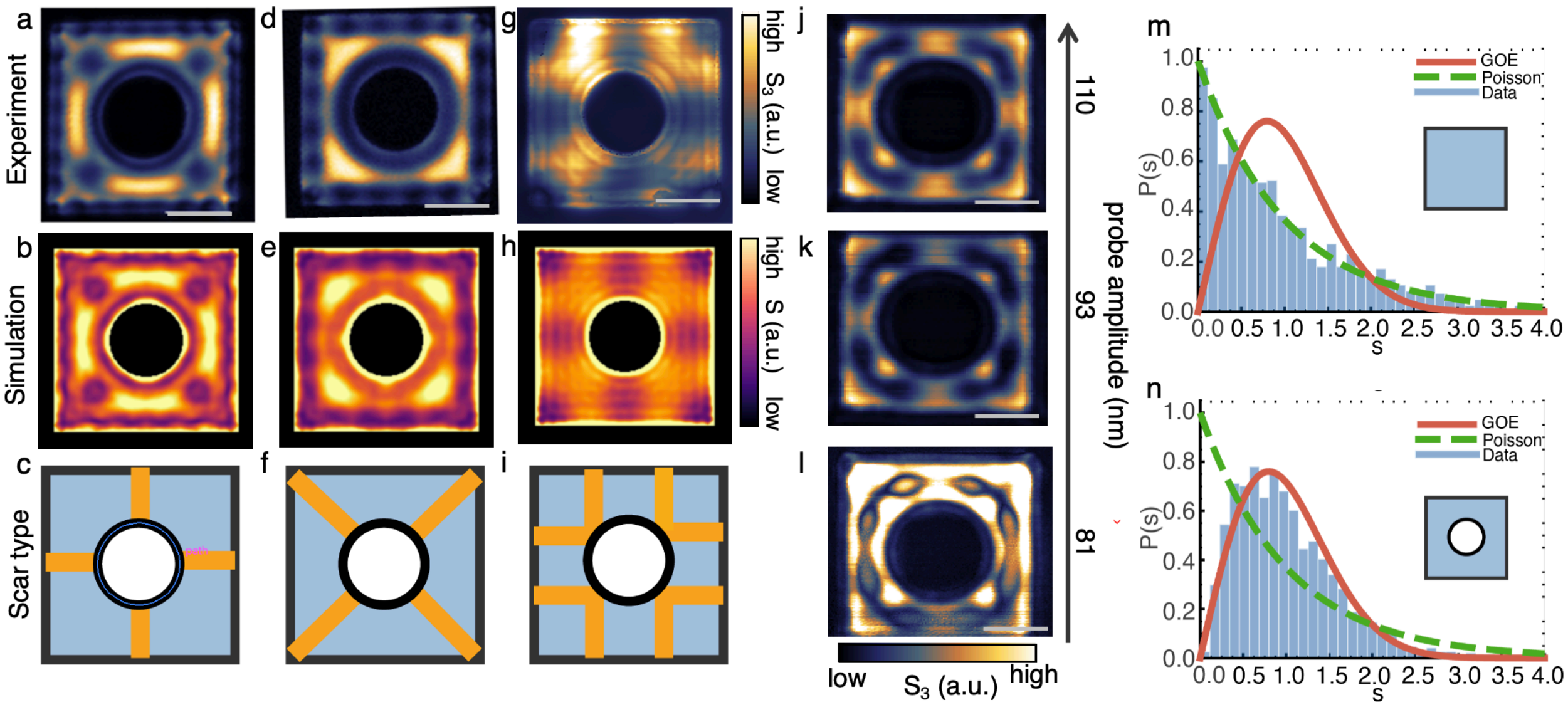

**Figure. 2 Polariton chaos in Sinai billiards** a.-i. Scarring effect of hBN phonon polaritons. The first row shows experimental SNOM data; the second row shows the corresponding Green's function simulations; and the third row illustrate the associated classical trajectories in orange lines. Scale bars, 1um. j.-l. SNOM data taken on the same 72-nm thick natural sample at ω = 1493 cm⁻¹, with varied probe amplitudes 110, 93 and 81 nm, respectively. m.-n. Calculated energy level spacing distributions of hBN polariton eigenmodes in a square billiard (m) and a Sinai billiard (n). The blue columns represent the normalized probability P(*s*), and x-axis shows the dimensionless nearest-neighbor energy level spacing, *s*. The dashed green and solid red curves represent theoretical fits to the Poisson and Gaussian orthogonal ensemble (GOE) distributions, respectively.

In addition to the single-color imaging experiments, we also performed broadband hyperspectral near-field imaging to compare nonchaotic edge modes and chaotic interior of the Sinai billiard. Figure 3a shows spectra taken along the billiard edge, where edge-mode phonon polaritons exhibit discrete energy levels corresponding to distinct mode numbers, consistent with an effectively one-dimensional integrable cavity. In contrast, spectra acquired inside the Sinai billiard (Fig. 3b) lose simple periodic real-space node structures, displaying irregular pattens with diagonal spectral features connecting different modes. Figure 3c-d are Green's function simulations for a-b, where key experimental features are qualitatively reproduced. While the hyperbolic phonon polaritons retain a robust quantum number for z-direction, the in-plane quantum numbers are governed by the chaotic cavity geometry.

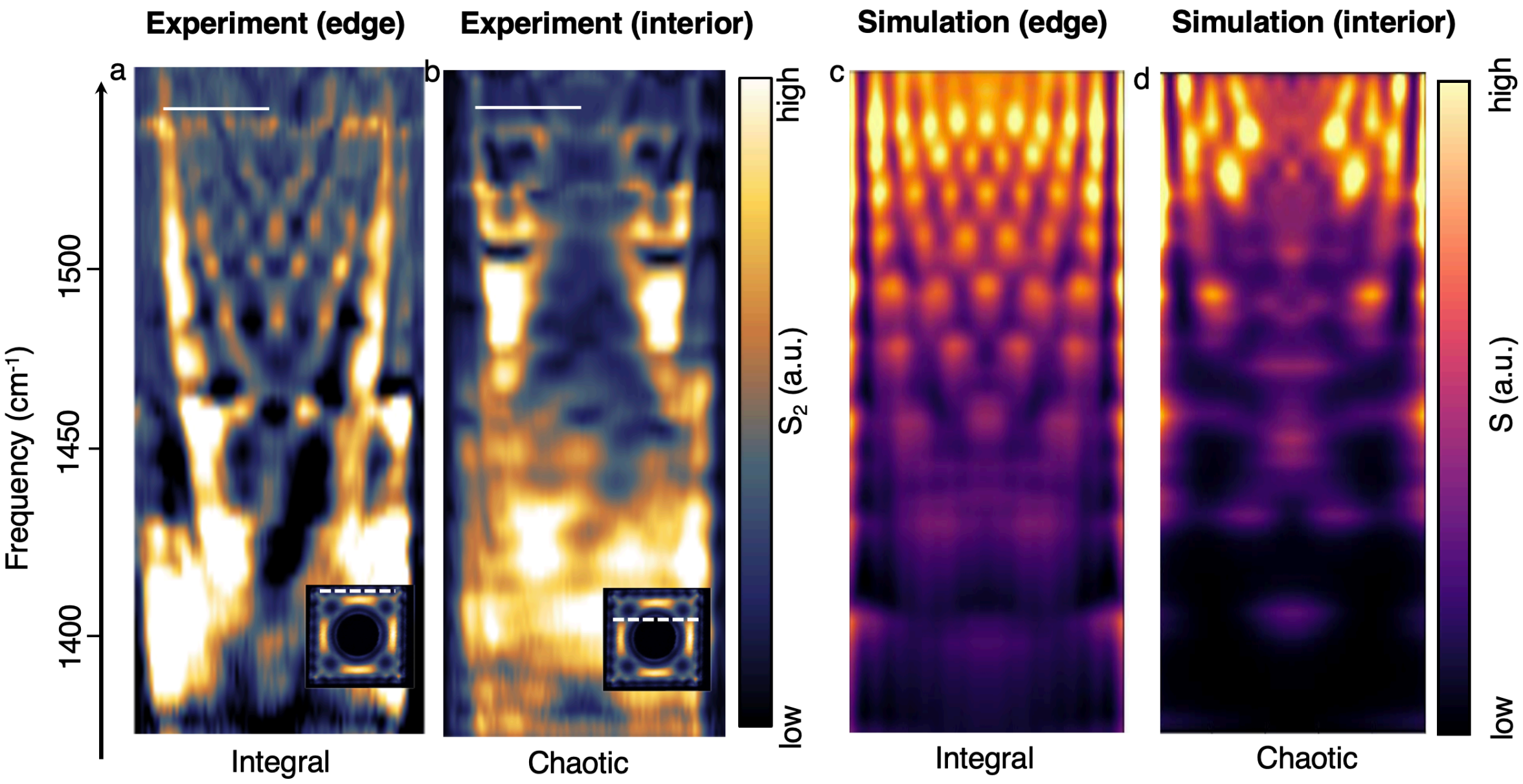

**Figure. 3 Chaos-induced modification of energy landscape of phonon polaritons.** a-b, Hyperspectral image taken along the edge and at the interior of a Sinai billiard of 100nm thickness and 5μm width. The locations where a-b are taken are marked by the dashed lines in the insets. Scale bar: 2μm. c-d. Simulated dispersions for a-b.

## Berry's conjecture and irregular boundaries

Berry's conjecture[30] points out that in the semiclassical limit, chaotic quantum energy eigenstates are close to Gaussian random fields exhibiting random interference patterns which are statistically isotropic. For polariton billiards, the conjecture implies that a chaotic cavity eigenmode is no longer composed of a few discrete momenta directions. Instead, polariton wavevectors should populate the allowed isofrequency surface in an approximately uniform manner. We observe hallmarks of Berry's random wave conjecture in complex billiards which is less clear in symmetric simpler geometries such as Sinai billiard.

Figures 4a–d present four cavity geometries with increasing boundary complexity, designed following Koch snowflake[31] inspired fractal boundaries. Starting from a hexagon corresponding to fractal order zero (Fig. 4a-c), each higher order boundary is generated by trisecting edges and inserting inward equilateral triangles (Fig. 4d-l). The three rows in Fig. 4 present the corresponding SNOM images, fast Fourier transforms (FFTs) of the data, and the simulation of FFTs. All experimental data shown here are acquired using a probe tapping amplitude of 106 nm with more probe amplitudes in SI. As the fractal order increases, the FFT images evolve from discrete momentum features (Fig. 4b-c) at $|k|$=25$\mu$m$^{-1}$ toward nearly isotropic rings (Fig.4k-l). The momentum of polaritons in a hexagon billiard exhibits six well-defined directions, reflecting the C6 symmetry. In the first-order convex polygon, these features become smeared along the isofrequency

contour (Fig. 4e and j), signaling the onset of wave randomness. With further increased fractal orders, the momentum distribution approaches an almost isotropic ring (Fig. 4k–l), indicating a progressively increasing uniformity of the polariton momentum directions as predicted by Berry's random-wave conjecture. We have also calculated the autocorrelation length for experimental data in Fig. 4 a, d, g and j, which are, 716.86 nm, 483.62 nm, 327.45 nm, and 254.95 nm respectively. As the order of boundary fractalness is increased, the convergence of autocorrelation length to the half of the free second-order polariton wavelengths (~500nm) serves as additional evidence[32] for the emergence of chaos. Further perturbation using different probe amplitudes and additional measurements on more random boundary geometries are presented in SI Sec. 3.4-3.5.

We highlight two perspectives from the experiments in Fig. 4. First, from the viewpoint of quantum-classical correspondence, polaritons are finite-size objects rather than point particles. Therefore, as discussed in Ref. [33], a convex polygon billiard such as a first-order Koch snowflake (Fig. 4d-f) remain nonchaotic in classical case regardless of the particle size. It is the quantum wave effect that allows chaotic behavior to emerge. Second, unlike the smooth convex scatterer in a Sinai billiard, the second- and third-order Koch snowflake geometries contain multiscale boundary features that can disrupt the local polariton phase. These features promote diffusive rather than specular reflection of ballistic polaritons. Thus, by increasing the complexity of irregular boundaries, we control the onset of chaotic polariton dynamics in a manner consistent with Berry's random-wave conjecture.

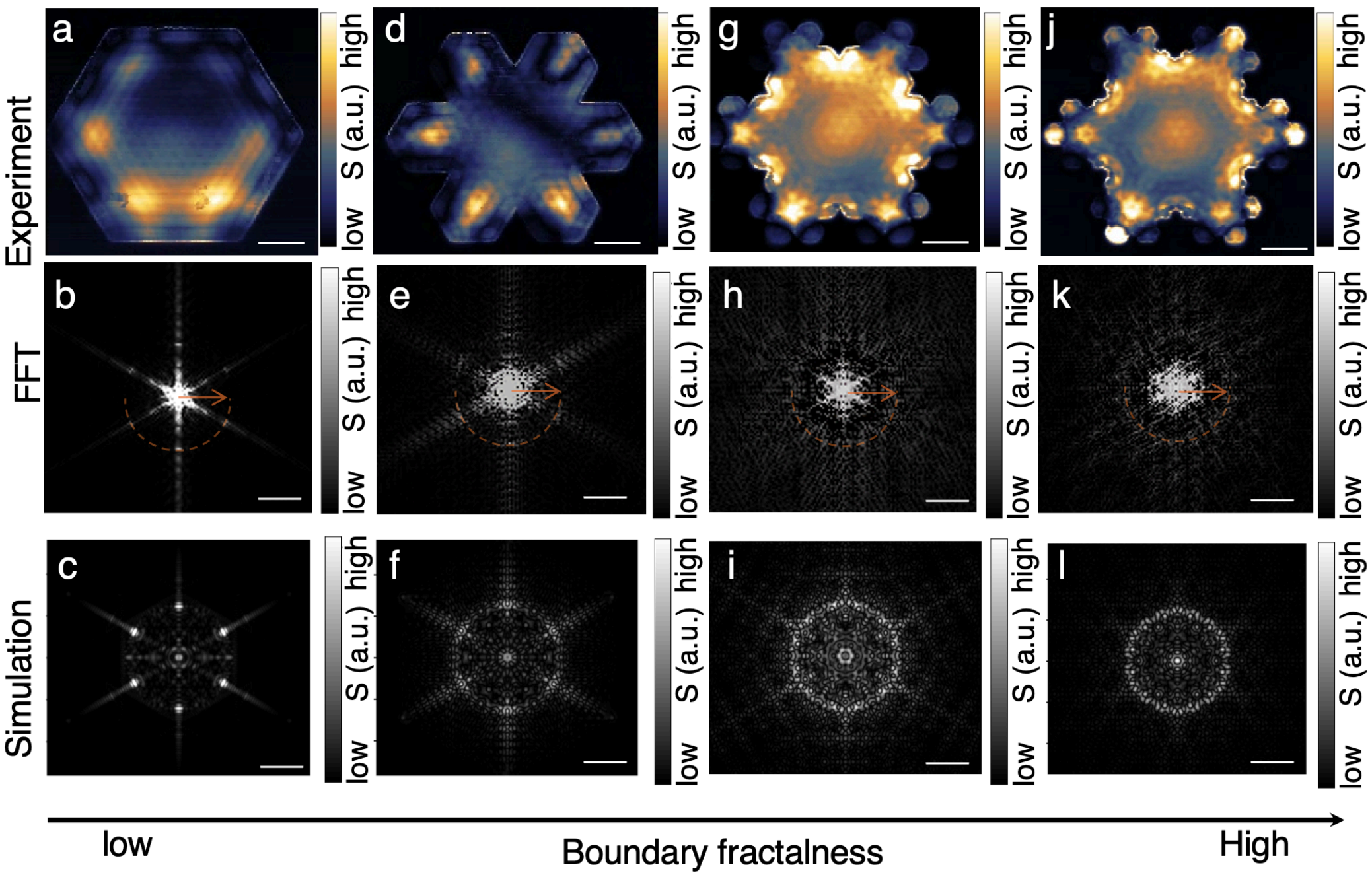

**Figure. 4. Observation of Berry's conjecture.** Polariton billiard imaging and corresponding FFT for (a-c) hexagon, (d-f) first-order, (g-i) second-order, and (j-l) third-

order fractal Koch snowflake cavities. The first row is experimental SNOM data, all taken at 1429 $cm^{-1}$ with a probe amplitude of 106 nm. Scale bars, 1 μm. The second row is the corresponding FFT analysis of the first row, where the orange arrows indicate polariton momentum $|k|=25\mu m^{-1}$ measured from the zone center, corresponding to the real-space fine features with a wavelength around 251 nm in first row. Scale bars, 20 $\mu m^{-1}$. The third row is simulated FFT results for the same billiard. Scale bars, 20 $\mu m^{-1}$. As the fractal order of the boundary increases, the spectral feature at $|k|=25\mu m^{-1}$evolves from discrete points into a continuous ring, conforming to Berry's conjecture of random waves.

**Summary and outlook**

Classical chaos theory was consolidated last century to describe dynamical systems that are sensitive to initial conditions[34,35]. By contrast, the quantum counterpart of chaos[36, 37,38] is still under development to date[39,40] and lack of a universal definition. Through systematic near-field imaging experiments and supported by theoretical modeling, we have observed multiple hallmarks of quantum chaos in hBN phonon polariton billiards, such as scarring, irregular cavity modes, and manifestation of Berry's conjecture. Specifically, we have addressed the key roles of geometry and boundary for inducing chaos in hBN phonon polariton billiards.

Quantum chaos is relevant to a variety of phenomena such as photon equilibration[41], entanglement[42], superconductivity[43], and information scrambling[44]. Looking forward, edge modes in chaotic and nonchaotic billiards offer a promising platform for engineering the energy dissipation from bulk modes. An interesting distinctive feature of polaritonic billiards is the coexistience of integrable edge modes and chaotic bulk. The former controls the dissipation of the latter which would be interesting target for future studies. With inter-cavity interactions[45], chaotic polariton dynamics may provide new engineering routes to photonic and phononic crystals. The framework established here is also generalizable to other light-matter hybrid excitations, including plasmonic, excitonic, and magnonic polaritons, opening new opportunities for exploring and manipulating light-matter interactions in complex and nonintegrable systems.

**Acknowledgement**

Y.D.,F.L.,Y.L., M.F. and D.N.B are supported by NSF ECCS-2529062 for cavity research, and in part by Julian Schwinger Foundation for polariton imaging. Y.D., D.S., C.R.D. and D.N.B. are supported by Programmable Quantum Materials, an Energy Frontier Research Center funded by the U.S. Department of Energy (DOE), Office of Science, Basic Energy Sciences (BES), under award no. DE-SC0019443. D.N.B. is a Moore Investigator in Quantum Materials EPIQS no. 9455. V.G. was supported by the U.S. Department of Energy, Office of Science Basic Energy Sciences under Award No. DE-SC0001911. V.G. is grateful to Sir Michael Berry for helpful comments on the random wave conjecture and classical-to-quantum correspondence.

[28] M.V. Berry, and M. Tabor, Level clustering in the regular spectrum. Proceedings of the Royal Society of London. Series A, Mathematical and Physical Sciences 356, 375–394 (1977).
[29] A. Backer, "Quantum Chaos in Billiards," in *Computing in Science & Engineering*, vol. 9, no. 3, pp. 60-64, May-June (2007)
[30] M. V. Berry, Regular and irregular semiclassical wavefunctions,Journal of Physics A 10, 2083 (1977).
[31] H. von Koch, “Une méthode géométrique élémentaire pour l’étude de certaines questions de la théorie des courbes planes”, Acta Mathematica 30, 145–174 (1906).
[32] J. Lin, Y. Zhuang, A. M. Graf, J. Keski-Rahkonen, and E. J. Heller, Shaping chaos in bilayer graphene cavities Arxiv, 2512, 10914 (2026)
[33] Efim B. Rozenbaum, L. A. Bunimovich, and V. Galitski, Early-Time Exponential Instabilities in Nonchaotic Quantum Systems, Phys. Rev. Lett. 125, 014101 (2020)

[34] H. G. Schuster and W. Just, Deterministic Chaos: An Introduction, 4th Ed. (2005).
[35] A. J. Lichtenberg and M. A. Lieberman, Regular and Chaotic Dynamics, 2nd ed. (Springer, 1992)
[36] F. Haake, Quantum Signatures of Chaos, 3rd ed. (Springer, 2010).
[37] M. V. Berry, “Quantum chaology,” Proceedings of the Royal Society A 413, 183 (1987).
[38] M. C. Gutzwiller, Chaos in Classical and Quantum Mechanics (Springer, 1990).
Operational definition of quantum chaos
[39] A. Kitaev, “ Hidden Correlations in the Hawking Radiation and Thermal Noise”, talk given at Fundamental Physics Prize Symposium, November 10, 2014. Stanford SITP seminars, November 11 and December 18, 2014. “A simple model of quantum holography”, KITP lectures (2015).
[40] J. Maldacena, S.H. Shenker, D. Stanford, A bound on chaos, JHEP 2016, 106 (2016)
[41] V. M. Bastidas, H. L. Nourse et al, Equilibration of noninteracting photons and quantum signatures of chaos, Physical Review B, 112, 13, 10 (2025)
[42] J. Wang, L. Wang et al. Many-body quantum chaos, localization, and multiphoton entanglement in optical synthetic frequency dimension, Physical Review Applied, 23, 3, 3 (2025)
[43] P. Jacquod, H. Schomerus et al. Quantum Andreev Map: A Paradigm of Quantum Chaos in Superconductivity, Physical Review Letters, 90, 20, 5 (2003)
[44] B.Swingle, Unscrambling the physics of out-of-time-order correlators. Nature Phys 14, 988–990 (2018).
[45] C-H Yi, HC Park & M J Park, Bloch theorem dictated wave chaos in microcavity crystals, Light: Science & Applications 12, 106 (2023)

Supplementary Materials for

***Signatures of quantum chaos in phonon-polariton billiards***

## 1. Experimental Methods

To prepare the samples, we exfoliate isotopically enriched hBN ($^{10}$B) crystals onto silicon substrates with a 285 nm $SiO_2$ layer. The flake thickness and surface cleanliness are prescreened by optical microscopy based on optical contrast, and subsequently verified by atomic force microscopy (AFM). The selected hBN flake is then transferred onto a silicon chip patterned with gold alignment markers using a PDMS (polydimethylsiloxane) stamp coated with a PPC (polypropylene carbonate) film. We use electron beam lithography to define the desired cavity geometry, followed by reactive ion etching with $SF_6$ to remove excess material. After fabrication, the sample is annealed at 400 °C in forming gas (5% $H_2$ and 95% $N_2$) for 10 minutes to remove residual PMMA and PPC.

Nano-imaging experiments were carried out at room temperature under ambient conditions using a commercial scattering-type scanning near-field optical microscope (s-SNOM) equipped with an FTIR module (Neaspec). Metallic AFM tips (NanoWorld) with an apex diameter of approximately 20 nm were operated in tapping mode, oscillating vertically with an amplitude of 60–120 nm. The AFM tips both locally launch polaritonic modes and outcouple the electromagnetic waves carrying information about the polariton fields. For single-color imaging experiments, tunable continuous-wave mid-infrared quantum cascade lasers (Daylight Solutions) were used. The single-color near-field imaging was performed using a pseudo-heterodyne interferometric scheme, in which the reference mirror was modulated at a frequency of several hundred hertz in addition to the tip oscillation frequency (~75 kHz). For hyperspectral imaging experiments, broadband mid-infrared pulses from difference frequency generation (DFG) utilized outputs of a dual channel optical parametric amplifier (OPA, Orpheus Twins). The amplifier was pumped by a 20 W, 1030 nm Yb:KGW laser (Pharos, Light Conversion) operating at 755 kHz. The idler was fixed at 1500 nm and the signal tuned to 1940 nm to produce broadband mid-infrared radiation. A piezo-driven stage was used to translate the reference mirror, allowing frequency-resolved measurements via a Michelson interferometer. The broadband near-field signals were detected using a self-homodyne interferometric scheme.

## 2. Supplemental Theory

In this section, we provide details on the theoretical approach we use to reproduce the experimental observations of phonon polaritons shown in Figs. 2-4 of the main text, including the spectral and the real space features. Since in the s-SNOM experiment the tip both launches (creates) and detects (annihilates) polaritons at approximately the same point in space, the relevant information is encoded in the polaritonic Green's function that represents the central object of our study. In order to numerically evaluate it, we approximate the Maxwell equations by a scalar Helmholtz equation for each polaritonic branch with the appropriate boundary conditions that account for the specific phase polaritons acquire when reflecting off edges. This is achieved by utilizing the open-source finite-element software NGSolve finite element library [1].

## 2.1 Boundary Condition

In hBN, the strong coupling between infrared photons and lattice optical phonons gives rise to hyperbolic phonon polaritons. While these quasiparticles are tightly confined within the sample, their behavior at material boundaries is generally non-trivial requiring solving the full Maxwell equations. In particular, it is known that a normally incident polariton reflects from the hBN-vacuum boundary with $-\pi/4$ phase [2]. Furthermore, this phase depends on the angle of incidence and frequency in a non-trivial way. In order to simulate such scattering, below we derive the corresponding boundary condition that models a generic scattering phase-dependence $\delta(\bf k)$ at an infinite planar boundary which is found to be non-local in the general case. We then approximate this boundary condition by expanding around the normal incidence.

We approximate the propagation of phonon polaritons in hBN slab using the scalar two-dimensional Helmholtz equation

$$\Delta\psi + k^2\psi = 0 \qquad \text{in } \Omega \tag{S1}$$

where $\psi$ is the polariton field, k is the effective wave number, and $\Omega \subset \mathbb{R}^2$ is a bounded, connected two-dimensional domain that defines the specific chaotic billiard geometry (e.g., a Sinai billiard). We note that since Eq. (S1) generally describes linearly dispersing quasiparticles, below we describe a nonlinear mapping from the Helmholtz eigenvalues obtained from Eq. (S1) to reproduce the more realistic phonon-polariton dispersion in hBN.

We take into account the angle-dependent reflection phase of phonon polaritons using a generalized boundary condition,

$$\partial_n\psi + f(\boldsymbol{k})\mathcal{T}_{\boldsymbol{k}}\psi = 0, \qquad \text{on } \partial\Omega \tag{S2}$$

where $\partial_n$ is the outward normal derivative, $\boldsymbol{k}$ is the wave vector, and the function

$$f(\boldsymbol{k}) = \tan\left(\frac{\delta(\boldsymbol{k})}{2}\right)$$

determines the reflection phase $\delta(\boldsymbol{k})$. We define $\mathcal{T}_{\boldsymbol{k}}$ as

$$\mathcal{T}_{\boldsymbol{k}} := \sqrt{|\boldsymbol{k}|^2 + \Delta_{\partial\Omega}}, \tag{S3}$$

with $\Delta_{\partial\Omega}$ the Laplace-Beltrami operator on the boundary $\partial\Omega$. We note that this boundary condition interpolates between Neumann ($f(\boldsymbol{k}) = 0$) and Dirichlet ($f(\boldsymbol{k}) \to \infty$) conditions.

We also note that strictly speaking our boundary condition Eq. (S2) is valid in the limit of a semi-infinite region. In order to demonstrate that, consider a plane polaritonic wave scattering off an hBN/vacuum boundary located at $x = 0$. Assuming hBN occupies region $x < 0$ and translational invariance along $y$ direction, we straightforwardly write

$$\psi(x,y) = (e^{ik_x x} + \mathcal{R}e^{-ik_x x})e^{ik_y y},$$

where $k_x$ and $k_y$ are propagation constants with $\mathbf{k}^2 = k_x^2 + k_y^2$ and $\mathcal{R}$ is the complex reflection coefficient. Applying the boundary condition Eq. (S2) to this ansatz, we find

$$\mathcal{R} = \frac{i+f(\boldsymbol{k})}{i-f(\boldsymbol{k})} = e^{-i\delta(\boldsymbol{k})}, \tag{S4}$$

where we used $\mathcal{T}_{\boldsymbol{k}} e^{ik_y y} = |k_x| e^{ik_y y}$. We thus find that our boundary condition Eq. (S2) correctly encodes an arbitrary phase shift acquired after scattering off a boundary.

While Eq. (S2) allows for the incorporation of the angle-dependent phase shift, the operator $\mathcal{T}_{\boldsymbol{k}}$ is highly nonlocal in real space, preventing direct implementation in confined geometries using standard finite element solvers, relying on sparse and local operators. To overcome this limitation, we approximate $\mathcal{T}_{\boldsymbol{k}}$ around normal incidence (i.e. effectively expanding in Laplace-Beltrami operator in Eq. (S3)) $\mathcal{T}_{\boldsymbol{k}} \approx k$. This reduces the non-local boundary condition to the following generalized Robin condition

$$\partial_n \psi + k\tan(\delta/2)\psi = 0, \tag{S5}$$

with $\delta$ being the normal incidence scattering phase shift. We note that the Robin coefficient in Eq. (S5) proportional to the eigenvalue of Helmholtz equation $k$. Therefore, the resulting set of equations represent an eigenvalue-dependent boundary-value problem. Numerically, we find solutions self-consistently for each eigenmode/eigenvalue pair.

Finally, As discussed above, the scattering in hBN is characterized by $\delta = -\pi/4$. In the following, we use Helmholtz equation Eq. (S1) supplemented with this boundary condition to find the spectrum of hyperbolic phonon-polaritons in confined geometries.

## 2.2 Simulation Framework

We model the s-SNOM measurement data using the local Green's function of phonon polaritons

$$G(\boldsymbol{r},\boldsymbol{r},\omega) = \sum_n \frac{\psi_n(\boldsymbol{r})^2}{(\omega+i\gamma_n)^2-\omega_n^2}. \tag{S6}$$

Here, $\psi_n$ denotes the eigenmodes obtained by solving the Helmholtz equation, Eq. (S1) with boundary condition Eq. (S5). The corresponding eigenfrequencies $\omega_n$ are extracted from the eigenvalues $k_n$ of Eq. (S1) via $\omega_n = f(k_n)$, where the function $f$ is obtained by fitting the phonon-polariton dispersion in hBN shown in Fig. 1b of the main text (see also Fig.S1). This mapping, adjusted for the different hBN slab thicknesses used in the experimental measurements, allows us to go beyond the linear dispersion imposed by the Helmholtz spectrum. Finally, we note that we also include a finite polariton decoherence rate, $\gamma_n = \omega_n/(2Q)$, where $Q$ is the quality factor.

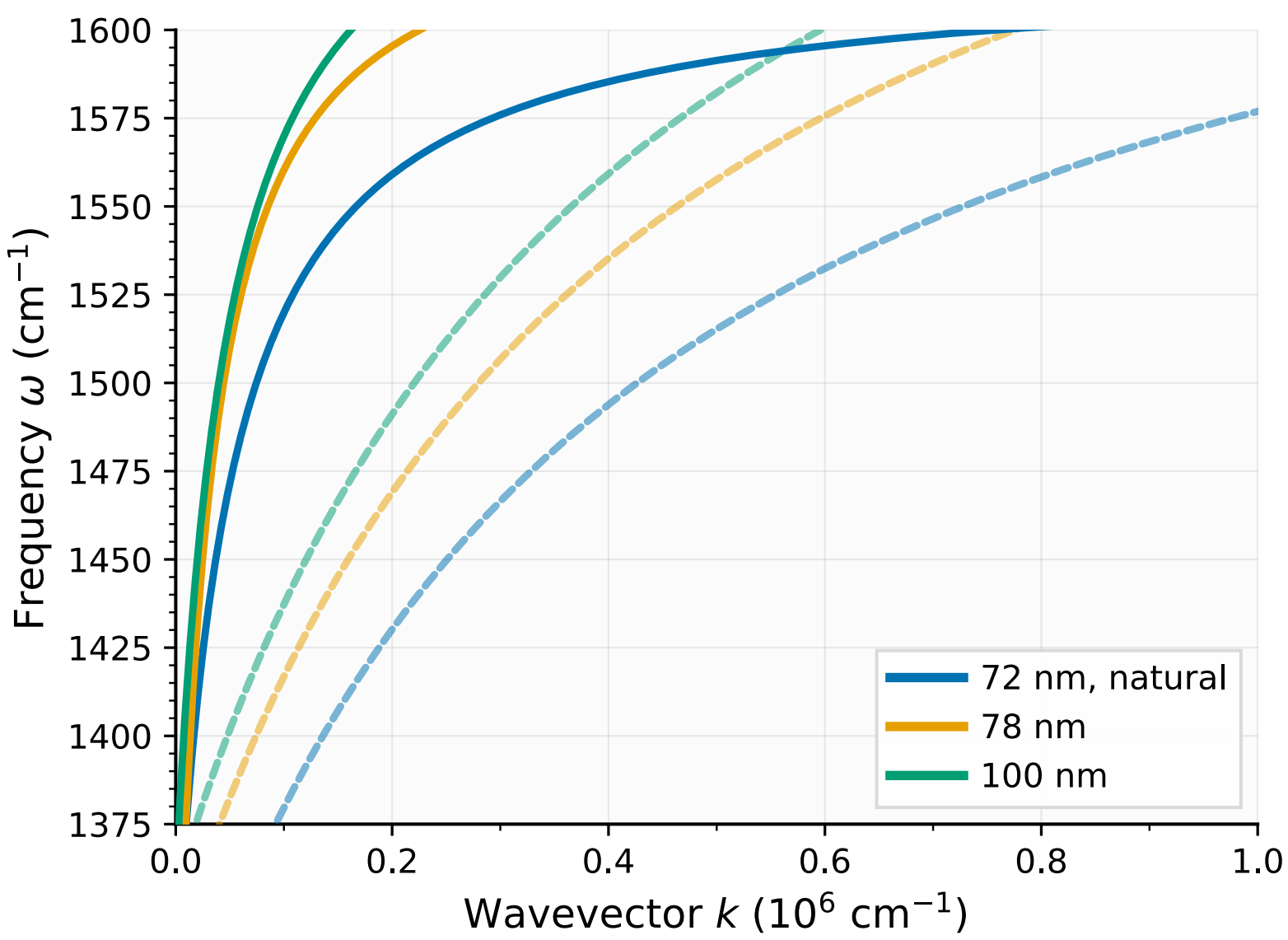

*Fig. S1. Polariton dispersion relations for various thickness of hBN slab used in simulations. Dahsed lines are the 2nd order polaritons.*

Our numerical results obtained within this approach are in good agreement with the experimental results, as can be seen in Fig. 2 and 3 of the main text. In particular, by incorporating the first 200 eigenmodes of Eq. (S1) with the boundary condition Eq. (S5), we recover essential spectral and spatial polaritonic features of the Sinai billiard.

## 2.3 Edge Modes

One of the key features of the polaritons observed in the experiments is the presence of *edge modes*. We define these edge modes as field configurations where the amplitude is significantly localized at the boundaries and decays exponentially into the bulk. As discussed in the main text (Fig. 3 of the main text), these modes effectively behave like one-dimensional, integrable systems, exhibiting discrete energy levels corresponding to distinct modes.

We now discuss how these edge modes are captured by our formalism. In particular, the emergence of these modes can be demonstrated by solving the Helmholtz equation Eq. (S1) in a semi-infinite slab, occupying $x \leq 0$. Using translational invariance along $y$ and assuming exponentially decaying solution along $x$ directions, we consider the following ansatz $\psi(x, y) = \psi_0 e^{\kappa_x x + i k_y y}$ with $\kappa_x, k_y \in \mathbb{R}$. Substituting this ansatz into Eq. (S5) at $x = 0$ yields

$$\kappa_x + k\tan\left(\frac{\delta}{2}\right) = 0. \tag{S6}$$

with $|{\bf {k}}|=\sqrt{k_y^2-\kappa_x^2}$. For consistency we need $\kappa_x > 0$ and $|k_y| > |\kappa_x|$. These conditions can be satisfied for the value $\delta = -\pi/4$ used in our simulations.

## 2.4 Berry Conjecture

According to Berry's random wave conjecture, high-energy eigenfunctions of classically chaotic systems behave locally, away from boundaries, as random superpositions of plane waves with fixed wavenumber and uniformly distributed propagation directions [3]. Consequently, their spectral weight in momentum space is isotropically distributed on a circle of radius $k$, yielding a characteristic circular pattern in Fourier space. In contrast, regular geometries, such as square or hexagonal billiards, exhibit highly structured discrete momentum distributions. In particular, as illustrated in Fig.S2, the Fourier transform of the solution $\psi$, for an integrable hexagon billiard exhibits discrete points reflecting six-fold symmetry, but as the geometry evolves into more complex fractals and non-integrable shapes, these discrete points transition into a continuous, isotropic ring.

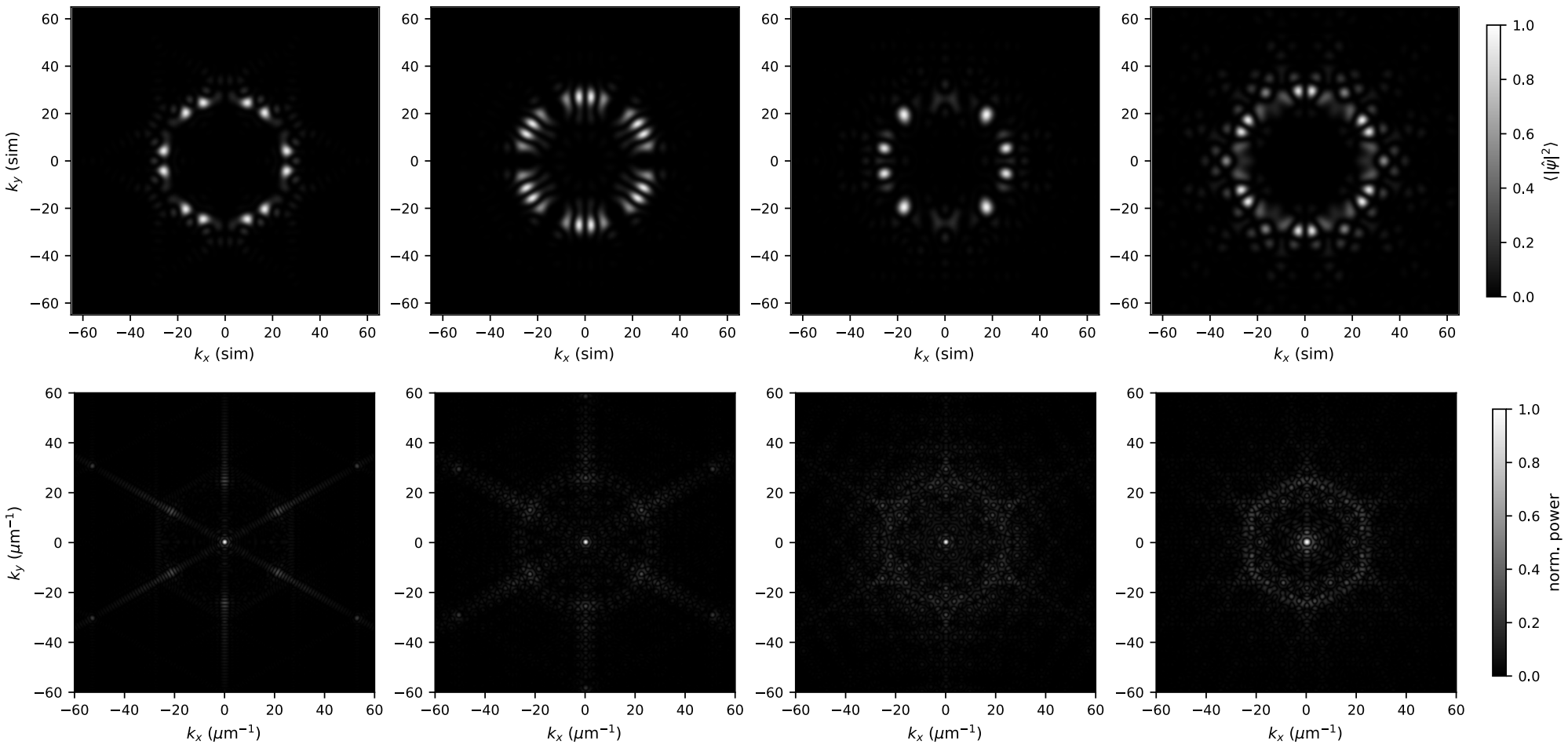

*Fig. S2 First row: Fourier transform of $\psi$. Second Row: Fourier transform of Green's function, $|G|$, with quality factor $Q = 100$.*

We note, however, that the experimental s-SNOM signal is proportional to the Green's function in Eq. (S6), and therefore probes the spatial structure of the eigenmodes through their local intensity. Moreover, in our analysis of the experimental data, we use the absolute value of the measured signal. The relevant information is thus encoded in the Fourier transform of the modal intensity,

$$\mathcal{I}_n(\boldsymbol{r}) \propto |\psi_n(\boldsymbol{r})|^2.$$

In momentum space this gives the convolution

$$\tilde{\mathcal{I}}_n(\boldsymbol{q}) \propto \int d^2\boldsymbol{k}\, \psi_n(\boldsymbol{k})\psi_n(\boldsymbol{q}-\boldsymbol{k}).$$

Wavefunctions in chaotic geometries $\psi_n$, obeying Berry's random wave conjecture, have momentum-space weight highly concentrated near the shell $|\boldsymbol{k}| = k_n$, with an approximately isotropic angular distribution. The convolution of two such isotropic shells has support for $|\boldsymbol{q}| \leq 2k_n$ and is enhanced near the outer boundary $|\boldsymbol{q}| = 2k_n$, apart from the contribution near $\boldsymbol{q} = 0$. Thus, in momentum space, the intensity exhibits an approximately isotropic disk with a pronounced outer ring at $q \simeq 2k_n$, consistent with our results illustrated in Fig. S2. In experiments on 68-nm-thick hBN, this $2k_n$ feature appears as a prominent isotropic ring-like structure for sufficiently chaotic shapes, providing a signature of chaotic phonon-polariton eigenmodes in these geometries.

### 2.5 Level statistics

To characterize the quantum chaotic nature of the phonon polaritons, we analyze the statistical distribution of the cavity's energy levels. In systems where the corresponding classical dynamics is integrable, the level statistics follow Poisson-like spectral statistics. Conversely, in classically chaotic systems, such as the Sinai billiard, the energy levels exhibit level repulsion characterized by Wigner-Dyson statistics [4,5].

To systematically compare the eigenmodes of our cavities with these universal predictions, we perform a spectral unfolding procedure. We map the raw eigenvalues to a dimensionless set of unfolded levels $\epsilon_n$ such that the mean level spacing is normalized to unity [6]. This unfolding relies on analyzing the cumulative spectral staircase staircase function,

$$N(k) = \sum_n \Theta\,(k - k_n),$$

which counts the total number of eigenmodes up to a given wavevector $k$. We separate the exact empirical staircase $N(k)$ into a smooth, continuous background $\bar{N}(k)$ and a fluctuating component $\delta N(k)$

$$N(k) = \bar{N}(k) + \delta N(k).$$

The leading smooth behavior for a two dimensional billiard of area $A$ is governed by Weyl's Law, $\bar{N}(k) = \frac{Ak^2}{4\pi}$. To account for finite-size boundary corrections, we calculate the residual fluctuations $\delta N(k) = N(k_n) - \bar{N}(k_n)$ and fit them with a high-degree polynomial $p(k)$ [7]. The unfolded eigenvalues $\epsilon_n$ are then obtained by

$$\epsilon_n = \bar{N}(k_n) + p(k_n).$$

From this normalized spectrum, we evaluate the nearest-neighbor spacings $s_n = \epsilon_{n+1} - \epsilon_n$. The probability density function of these spacings, $P(s)$, serves as a primary diagnostic for wave chaos.

As demonstrated in Fig. 2m,n of the main text, regular integrable geometries (such as the square billiard) yield Poisson distribution, $P_{\text{Poisson}}(s) = e^{-s}$ . In contrast, chaotic geometries (such as the Sinai billiard) exhibit modeled by the GOE Wigner surmise

$$P_{\text{GOE}}(s) = \frac{\pi}{2} s \exp\left(-\frac{\pi}{4} s^2\right).$$

### 2.6 Polartion in a box

Without formal treatment [8] for the quantization of dispersive phonon polariton modes, one can use the following simplified model to understand the in-plane quantization behavior of integrable polariton cavities. Consider the Helmholtz eigenproblem in a 1D cavity,

$$\frac{d^2}{dx^2}E_m(x) + k_m^2 E_m(x) = 0$$

Boundary conditions leads to eigenwave solutions or quantization of polariton momentum dispersion. Take Dirichlet condition as a simple example, $E(0) = E(L) = 0$ gives eigenwaves $E_m(x) = E_0 \sin\left(\frac{m\pi x}{L}\right)$, where $m$ is the eigenmode index and quantum number. Therefore, the allowed wavevectors are $k_m = \frac{m\pi}{L}, m = 1,2,3, \ldots$ We can formally separate the contribution from dispersive media (matter) and confined electromagnetic waves (light) in the weak-coupling regime, where $n_{\text{eff}}(k_p)$ is effective refraction index and $k_m$ is polariton eigen momentum, respectively.

$$\omega_m = \frac{c}{n_{\text{eff}}(k_p)} k_m = \frac{m\pi c}{n_{\text{eff}}(k_p)L}, \quad m = 1,2,3, \ldots$$

Here $n_{\text{eff}}(k_p)$ inherit the original free polariton dispersion, while $k_m$ determines the quantization of momentum. Experimentally, the energy spread of each quantized momentum or the gap between each separated dispersion section is related to the intrinsic cavity property. For chaotic billiards, both the regular momentum quantization and the discrete dispersion branches are strongly modified, as illustrated in the main text (Fig. 3).

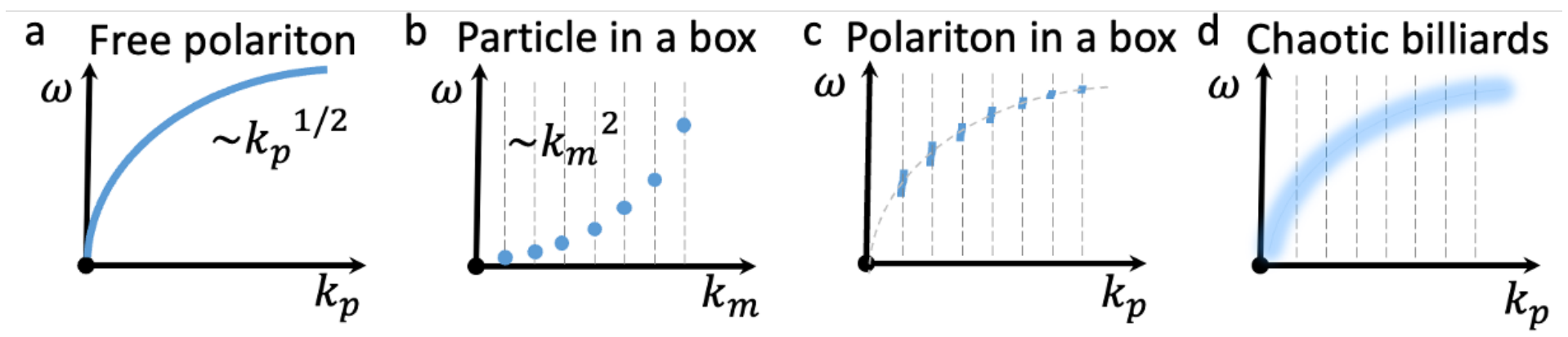

*Fig. S3 Comparison of dispersions for free polariton, particle in a box, and polariton in a box*

## 3. Supplemental Discussions

### 3.1 Enhancement of higher order polaritons in a cavity

While Purcell effect originally describes the enhancement of spontaneous emission [9], it can be generalized to absorptive and dispersive systems [10]. In scattering-type SNOM experiments, although phonon polaritons are probed through s-SNOM signals which are typically understood as local reflectance and absorbance [11], the metallic tip also acts as a nanoscale dipolar emitter and receiver whose coupling rate to phonon-polaritonic modes is governed by the local polaritonic local density of states (LDOS). Therefore, it is legit to borrow the Purcell language to describe the polaritonic enhancement in cavities [12,13].

We have observed that higher order hBN phonon polaritons are more prominent in cavities than first-order ones, in stark contrast with the free-propagation case. Here we qualitatively describe the cavity enhancement of different orders' polariton LDOS using Purcell-like factor, $F_P \propto \frac{Q}{V_{\mathrm{eff}}}$, where $Q$ is the intrinsic quality factor of the phonon polaritons and $V_{\mathrm{eff}}$ is the mode volume. Within the same hBN slab of thickness $\delta_z$, higher-order phonon polaritons with much larger in-plane momentum ($k_{//}$) possess smaller mode volume $V_{\mathrm{eff}} = \lambda_{//}^2 \delta_z$. Therefore, the cavity enhancement factor $F_P \propto Q k_{//}^2 \frac{1}{\delta_z}$ scaling quadratically with $k_{//}$ is larger for higher-order polaritons. In the calculated free polariton dispersion in Fig. 1, $k_{//,2nd} = 7.7 k_{//,1st}$. Consequently, even if the intrinsic quality factor of the second or higher order polariton is 50 times smaller than the first order polariton, with Purcell-like cavity enhancement, the observable LDOS for second or higher order polaritons can be larger than first order, as demonstrated in Fig.1 in the small cavities. An alternative perspective is that the higher order polaritons are much slower than first-order ones, resulting in longer interaction time with the hyperbolic cavity per round trip. In addition to the mode volume and interaction time consideration discussed above, the role of boundary condition is also non-negligible. For different orders of hyperbolic polaritons, excited by the same energy within the same slab, the effective wave reflection phases are different, leading to different energy dissipation at the cavity boundaries.

## 3.2 Impact of Sinai billiard geometry on scarring

As discussed in the main text, the formation of scarred eigenmodes requires the intrinsic polariton propagation length to exceed the characteristic cavity dimensions. With an approximately constant quality factor, the propagation length is primarily controlled by the polariton wavelength, and therefore by the excitation energy. Figures S4a to f compare devices measured at identical excitation frequencies (1465 to 1500 cm$^{-1}$) and identical hBN thickness, ensuring the same free polariton wavelength within each image. Under these fixed conditions, the billiard size and the relative size of the central obstacle strongly modify the spatial structure of the eigenmodes. The first number in parentheses denotes the ratio of the central perimeter to the cavity width (0.3, 0.4, 0.5), while the second indicates the cavity width (3, 4, 5 $\mu$m). Figures 4f to j present the same datasets over a larger field of view with an expanded intensity scale to emphasize enhanced scarring features.

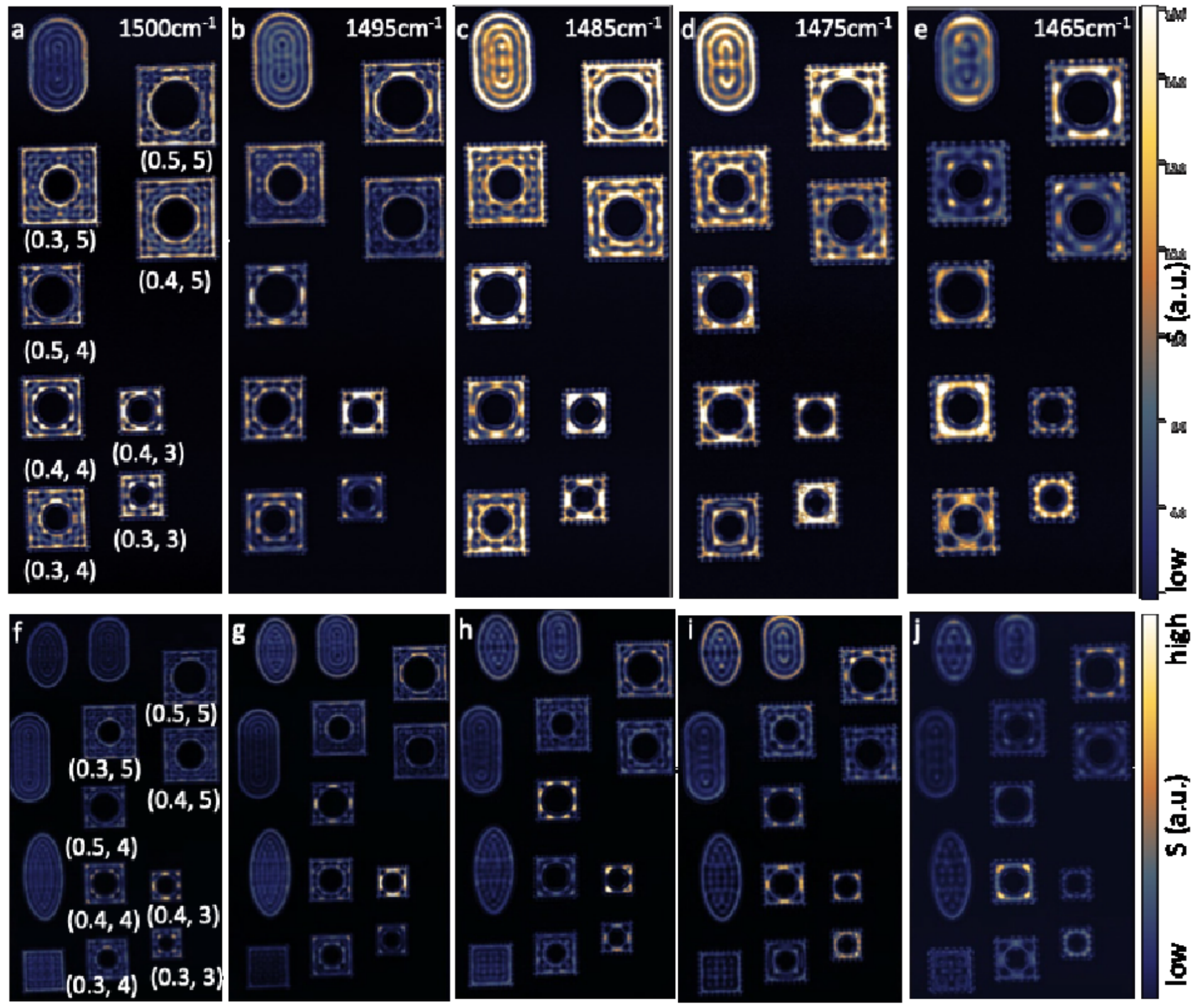

*Fig. S4 Geometric dependence of phonon polariton eigenmodes in Sinai billiards. Near field imaging maps are acquired for devices with the same thickness, 54 nm while varying the geometry and excitation energy. The first number in parentheses denotes the ratio of the central perimeter to the billiard width (0.3, 0.4, 0.5), and the second number indicates the cavity width (3, 4, 5 μm). Representative images are shown at six excitation frequencies, (a) 1465, (b) 1475, (c) 1485, (d) 1495, (e) 1500 cm$^{-1}$. (f–j) The same data as (a–e) with zoomed-out area and color scale to show the brightness of scars.*

Three characteristic classes of scarred modes are observed, whose prominence evolves systematically with polariton wavelength, cavity size, and geometric ratio. In general, smaller cavities and larger central obstacles favor stronger scar formation. At shorter wavelengths (Fig. S4a, f), cross shaped scars are most pronounced for the (0.4, 3) and then (0.3, 3) billiards rather than larger ones. As the wavelength increases (Fig. S4b, g), similar patterns emerge in the (0.5, 4) cavities due to its large obstacle ratio. At intermediate wavelengths (Fig. S4c, h), cross shaped scars transform into whispering gallery modes in (0.5, 4) cavities and into bright corner localized states in (0.4, 3). With further polariton wavelength increase (Fig. S4d), cross shaped scars dominate in (0.4, 4), while whispering gallery modes develop in both (0.5, 4) and (0.3, 3). The corner bright state for (0.4, 3) also become whispering gallery modes in Fig. S4e. For the largest cavities (5 μm), scarring becomes prominent only at the longest wavelengths 1465 and 1475 cm$^{-1}$.

## 3.3 Impact of quality factor on scar formation

In the main text, Fig. 2 shows that assuming the same quality factor, a larger wavelength leads to a longer propagation length and more prominent scarring. Here, we further demonstrate that with the same polariton wavelength, a higher quality factor leads to a longer propagation length and, therefore, prominent scarring, as an extension to the nonchaotic regime in Fig. 2c. Fig. 5 reveal that scarring effect begins to appear as the quality factor increases to approximately 100, around which the polariton propagation length becomes comparable to the billiard width, 4 $\mu$m.

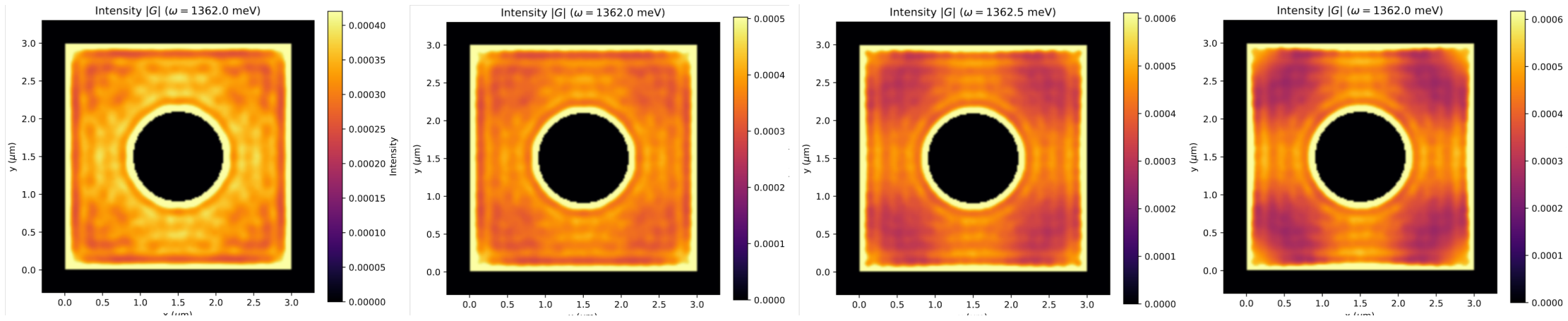

*Fig. S5 Simulated quality factor dependent scar appearance for* $Q = 50{,}100{,}300$ *and* $500$.

## 3.4 Control group and other shapes of billiards for Berry's conjecture tests

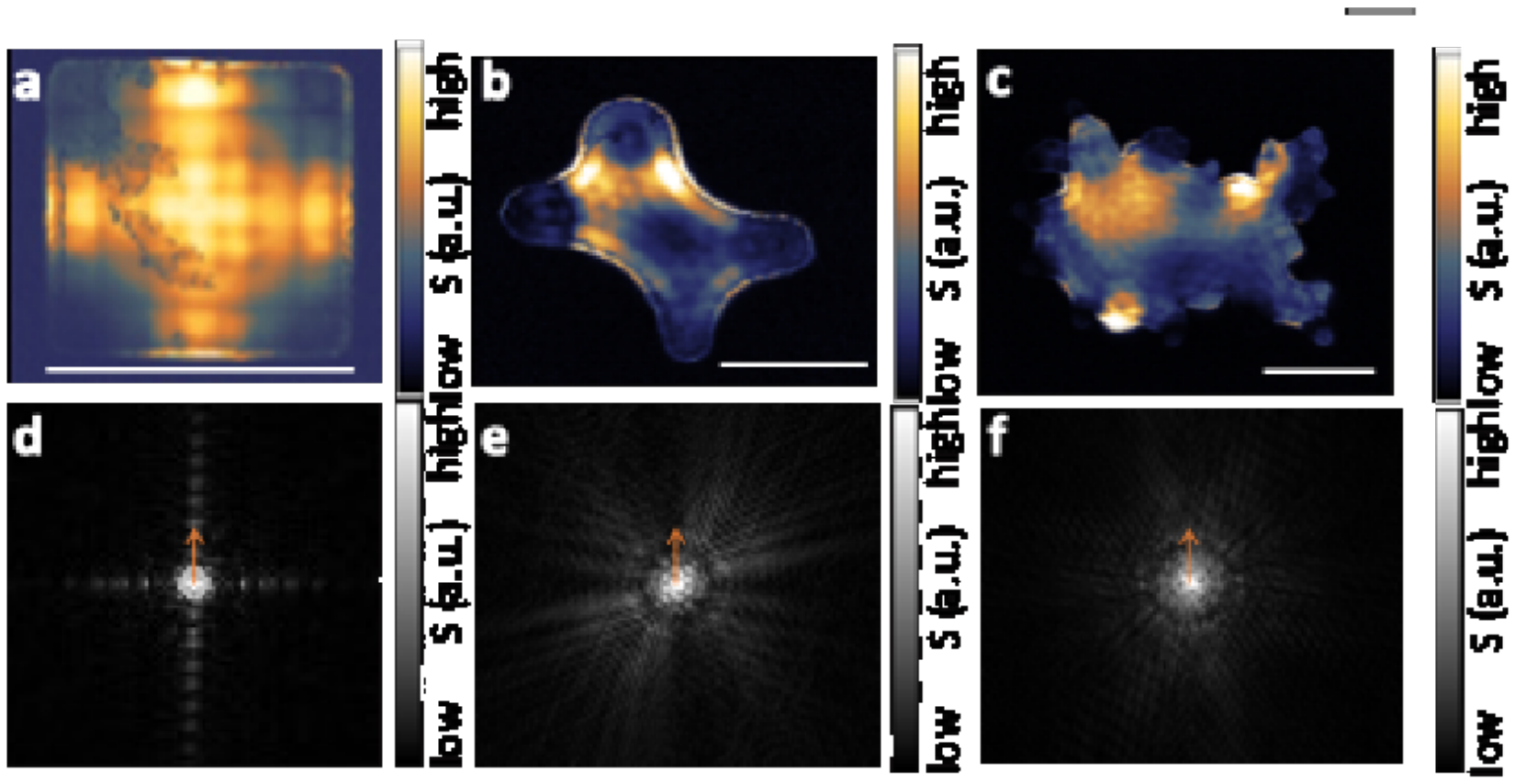

*Fig. S6 Control group of a square polariton billiard, fourth order SNOM image taken on 68 nm-thick hB*$^{10}$*N with 1429 cm*$^{-1}$ *laser. The dominating fine grid features correspond to second and higher order phonon polaritons. The brightened profile originates from the first-order polariton cavity mode. b. An asymmetric bone-shape billiard with four convex boundaries of four different curvatures therefore forming a chaotic billiard. c. A billiard with completely random shape and therefore chaotic. Scale bars, 2 μm. d–f. Corresponding FFT maps of a–c. The orange arrows mark the momenta of the second-order polariton modes with length* $|k| = 4\ \mu\mathrm{m}^{-1}$.

The data in Fig. S6 are all taken with the same material and excitation conditions, on differently shaped 68 nm-thick hB$^{10}$N billiards excited with 1428 cm$^{-1}$ laser. The polariton field in Fig. S6a follows the geometric shape of a square, showing corresponding four-fold symmetry in FFT. This is similar to the behavior of the hexagon billiard and corresponding six-fold symmetry as polariton momentum distribution inherits the integrable shape of billiards. However, for non-integrable shapes, such as a bone with four different curvatures (Fig.S6b), and a random map (Fig. S6c), the polariton momentum distribution, while keeping the same magnitude, can access random directions which manifested as an isotropic ring (Fig. S6e–f). This behavior further confirms the universality of Berry's random wave conjecture, in addition to the high-order Koch flakes studied in the main text.

## 3.5 Impact of probe amplitude on chaos

In Fig. 2, we demonstrated that the probe amplitude can tune the formation of first-order polariton mode patterns in Sinai billiards. Here we show that this perturbation effect from the probe applies to higher-order polariton modes and other billiard shapes as well. Fig. S7a and b show the SNOM images of higher order polariton mode patterns taken at 68 nm-thick $hB^{10}N$ with 1429 $cm^{-1}$ laser on second and third order Koch snowflake billiards. They are with a smaller probe amplitude, 66 nm, rather than in Fig. S7c and d and main text. All other parameters are kept the same. From a–d, it is evident that the brightened scarring modes around the convex edges are enhanced with larger probe amplitude. In the meantime, larger probe amplitudes lead to a more isotropic and randomized momentum distribution.

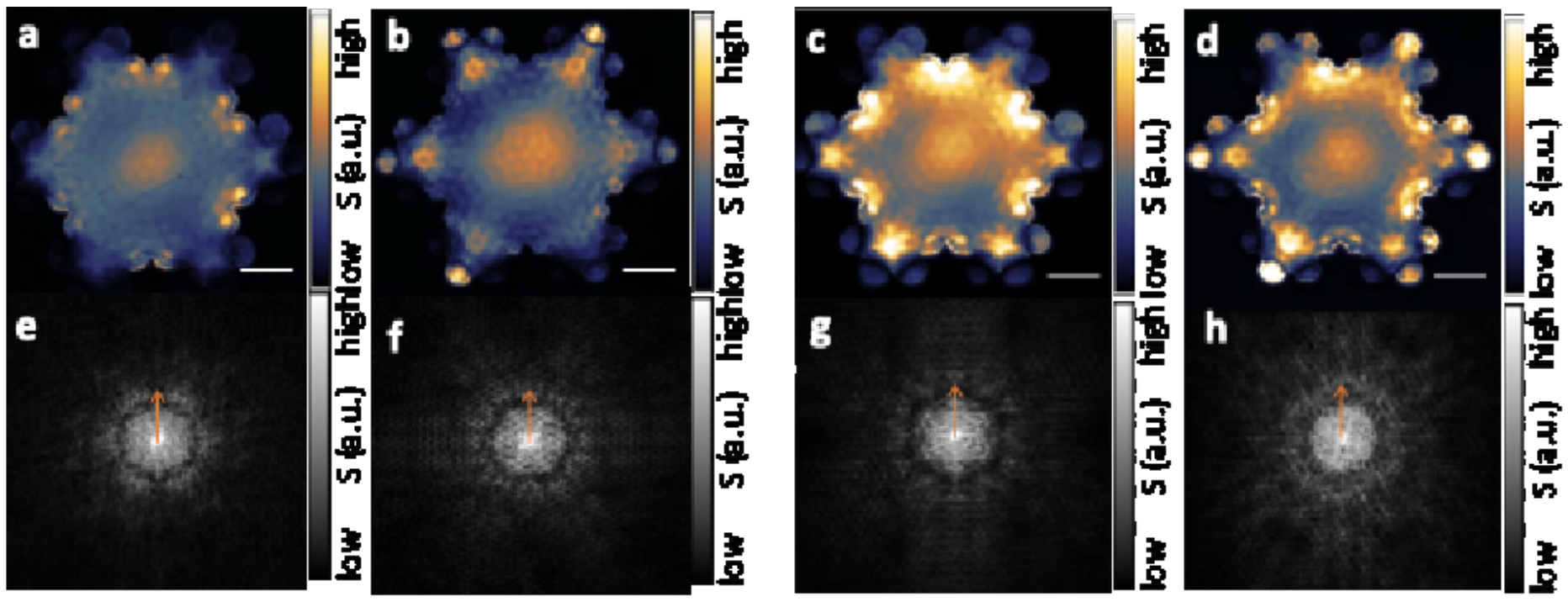

*Fig. S7 SNOM images of polaritons in second (a, c) and third (b, d) order Koch snowflakes taken on 68 nm-thick $hB^{10}N$ with 1429 $cm^{-1}$ laser with different probe amplitudes: a–b, 66 nm, c–d, 106 nm. e–f, corresponding FFT maps of a–c. The orange arrows mark the momenta of the second-order polariton modes with length $|k| = 4\ \mu\mathrm{m}^{-1}$.*

In the main text, we have shown that the sensitivity of polariton momentum distributions to the probe amplitude serves as one of the signatures of the system's chaotic nature. Here we explain this parameter in more detail. The probe amplitude is a tapping-mode atomic force microscopy (AFM) parameter that corresponds to the average oscillation height of the metallic tip above the sample surface. As shown in Fig. S8d, the probe amplitude is typically on the order of one hundred nanometers, which is one order of magnitude smaller than the near-field polariton wavelength and two orders of magnitude smaller than the far-field excitation wavelength.

The mechanism by which probe amplitude alters the detected polariton modes in chaotic billiards can be tentatively understood as follows. First, for the polariton excitation process, different tip heights correspond to different initial conditions which modify the launched near-field wave profile and therefore result in different electromagnetic field distribution. Second, for signal detection process, tip collecting signal at slightly different heights is essentially maping the field intensity at a 2D slice of the 3D mode profiles which is beyond our simulation framework.

To address the tip height effect within 2D Greens function framework, we tentatively consider the role of tip as a dipole with different distances above the 2D sample in Fig. S8. As established in section 2, with weak probe-sample coupling, SNOM signal is primarily governed by the intrinsic electromagnetic response of the sample where the polariton interference patterns are quasi-static responses associated with the sample Green's function. We can further project pre-calculated sample responses onto tip excitation and collection vectors derived from a dipole model, and spatially evaluate the results to emulate a raster-scan measurement. This formulation therefore describes the role of the tip as an effective weighting of intrinsic sample responses, which can lead to different polariton mode patterns, as demonstrated in Fig. S8.

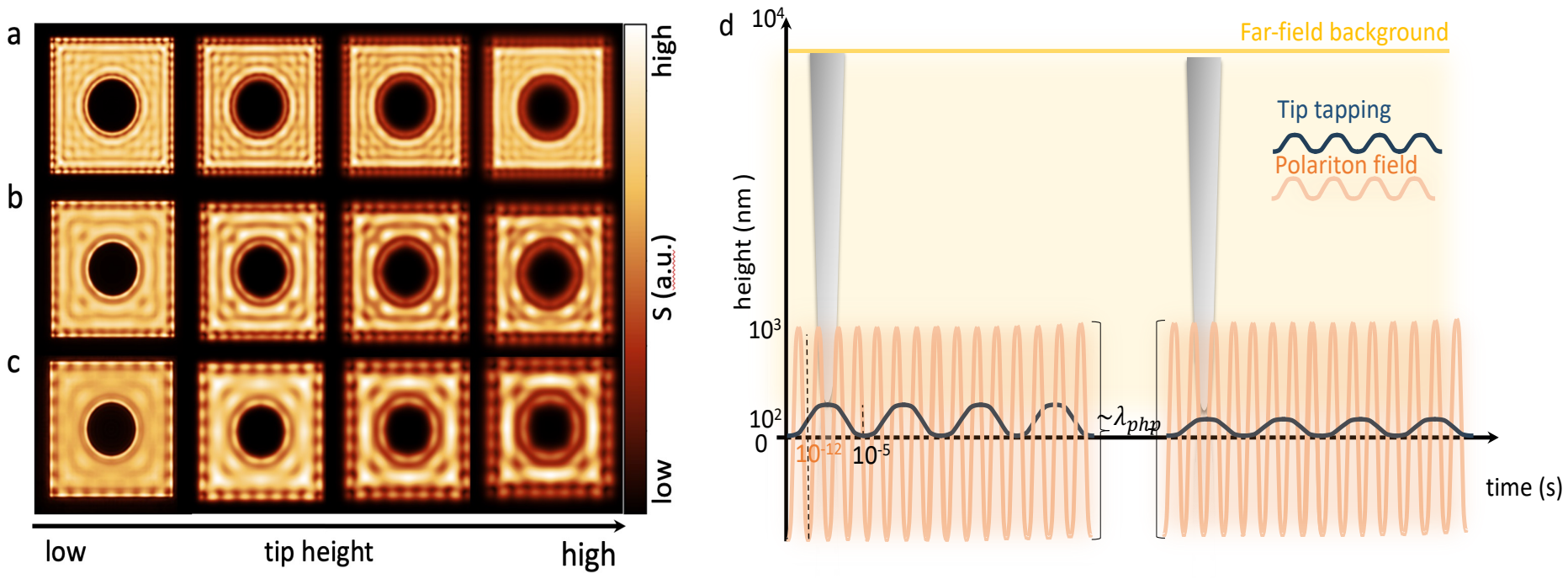

*Fig. S8 a–c. Simulated tip height impact on detected mode shape. d. Illustration of the length scale for probe amplitude, polariton field (first-order), and far-field background.*

## 3.6 Diverse scarring eigenmodes

As shown in the main text and Fig. S5, scarring effect manifests itself in polariton billiard problem as increased LDOS along underlying periodic classical orbits. Here (Fig.S9) we present a few more representative cases in addition to Fig. 2 in the main text to show the typical manifestation of cross-shape and diagonal scarring [14] in polariton billiards.

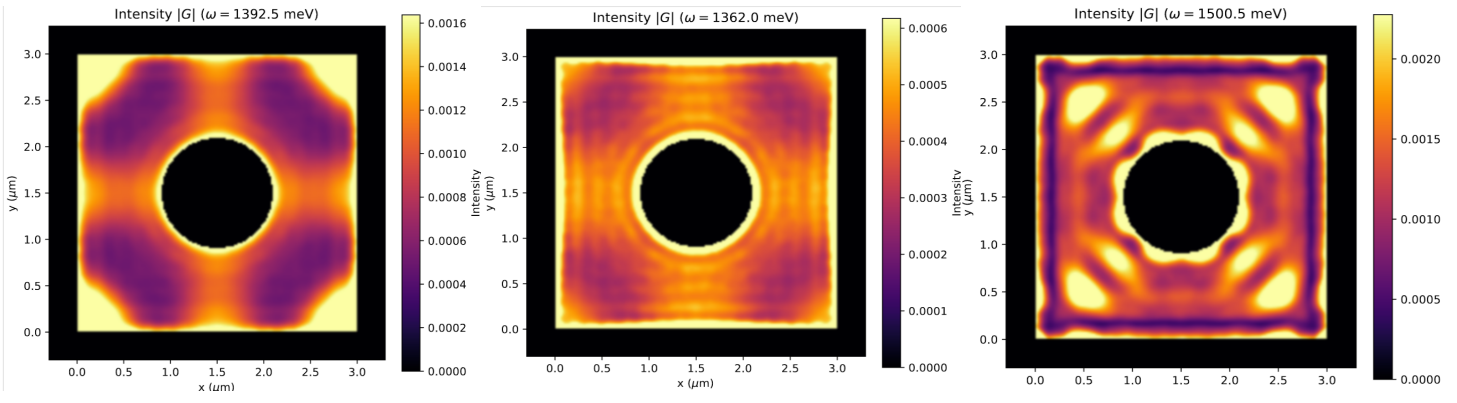

*Fig. S9 Other possible scars from simulation*